\documentclass{jfm}
\usepackage{graphicx,natbib,color,amssymb,amsmath}
\title[Effect of the number of vortices on the torque scaling in Taylor-Couette flow]{Effect of the number of vortices on the torque scaling in Taylor-Couette flow}
\author[B. Mart\'{i}nez-Arias and others]%
{B. Mart\'{i}nez-Arias,\ns
J. Peixinho\thanks{Email address for correspondence: jorge.peixinho@univ-lehavre.fr},\ns
O. Crumeyrolle\ns 
and I. Mutabazi}
\affiliation{Laboratoire Ondes et Milieux Complexes, Universit\'e du Havre \& CNRS UMR $6294$, \\
$53$ rue de Prony, $76600$ Le Havre, France\\[\affilskip]}
\date{\today}
\begin{document}
\maketitle
\begin{abstract}
Torque measurements in Taylor-Couette flow, with large radius ratio and large aspect ratio, over a range of velocities up to a Reynolds number of 24 000 are presented. Following a specific procedure, nine states with distinct number of vortices along the axis were found and the aspect ratio of the vortices were measured. The relationship between the speed and the torque for a given number of vortices is reported. In the turbulent Taylor vortex flow regime, at relatively high Reynolds number, a change in behaviour is observed corresponding to intersections of the torque-speed curves for different states. Before each intersection, the torque for a state with larger number of vortices is higher. After each intersection, the torque for a state with larger number of vortices is lower. The exponent, from the scaling laws of the torque, always depends on the aspect ratio of the vortices. When the Reynolds number is rescaled using the mean aspect ratio of the vortices, only a partial collapse of the exponent data is found.
\end{abstract}



\section{Introduction}
Since the works by \cite{Mallock1888}, \cite{Couette1890}, \cite{Taylor1936} and \cite{Wendt1933}, there was theoretical and experimental interest in the torque from liquids confined between two concentric differentially rotating cylinders.

Following the work by \cite{Taylor1923}, important contributions in the understanding of this flow instability have been made by many authors, including \cite{Stuart1958}, \cite{Donnelly1960}, \cite{Coles1965}, \cite{Snyder1969}, \cite{Gollup1975}, \cite{Barcilon1979}, \cite{Koschmieder1979}, \cite{Mullin1980}, \cite{DiPrima1981}, \cite{Benjamin1982}, \cite{Nakabayashi1982}, \cite{Riecke1986}, \cite{Cliffe1992}, \cite{Lathrop1992}, \cite{Lewis1999}, \cite{Takeda1999}, \cite{Xiao2002}, \cite{Czarny2003}, \cite{Abshagen2005}, \cite{Lim2004}, \cite{Racina2006}, \cite{Burin2010}, \cite{Dutcher2009} and many others. Because of these successes, the Taylor-Couette system is used to investigate the turbulence scaling laws, especially the behaviour of dimensionless torque acting on the rotating cylinder as a function of the Reynolds number.

From a practical point of view, \cite{Wendt1933} and \cite{Donnelly1960} provided scaling laws of the dimensionless torque, $G$, as a function of Reynolds number, {\it Re},  {\it i.e.} $G\propto Re^\alpha$ with an exponent, $\alpha$, that depends on {\it Re} and other parameters. More recent studies \citep{Lathrop1992,Lewis1999,Dubrulle2005,Ravelet2010,Paoletti2011,vanGils2011} provided additional data on torque and were concerned with fairly larger Reynolds numbers. 

\cite{Eckhardt2007} proposed analogies between turbulence in Rayleigh-B\'{e}nard convection and Taylor-Couette flow with emphasis on the evolution of the exponent. An argument was put forward to investigate turbulent Taylor-Couette flow because it has a stronger driving than Rayleigh-B\'{e}nard flow and should allow access to the ultimate turbulent regime \citep{Huisman2012,Ostilla2014}. This unifying theory was tested using the data of \cite{Lathrop1992} and \cite{Lewis1999}. They found a good agreement for a range of Reynolds above $10^4$.

From a numerical point of view, several groups \citep{Coughlin1996,Batten2002,Bilson2007,Pirro2008,Brauckmann2013,Ostilla2013,Ostilla2014} were able to simulate numerically turbulent Taylor-Couette flow up to relatively high {\it Re}. Most of these investigations \citep{Brauckmann2013,Ostilla2013,Ostilla2014} use periodic boundary conditions with relatively short calculation domains. \cite{Ostilla2013} presented the effect of three and four vortex pairs on the dimensionless transport suggesting that the larger number of vortex pairs induces an increase in torque. \cite{Brauckmann2013} investigated the effect of the vortex size on the torque and found a maximum of torque for vortices of axial wavelength of 1.93 times  the gap width for $Re=5000$. Hence, the Taylor-Couette system offers an opportunity to vary the number of vortices for a fixed value of the aspect ratio and therefore to modify the vortices shape and the shear between two adjacent vortices. Yet, all the recent studies do not explicitly mention the number of cells, although both \cite{Lathrop1992} and \cite{Lewis1999} provided torque data for eight and ten vortex states. 

In this article, we elucidate the effect of the number of vortices on the torque-speed relationship in a system containing up to nine states. This article is organised as follows. A description of the experimental apparatus is given in \S2 together with a detailed discussion of the protocols to obtain the different number of vortices. This is followed, in \S3, by the results on the torque data which are analysed. Finally, we draw some conclusions in \S4.

\section{Experimental setup and procedure}
\subsection{Experimental setup}

\begin{figure}
	\centering
	\includegraphics[width=0.6\columnwidth]{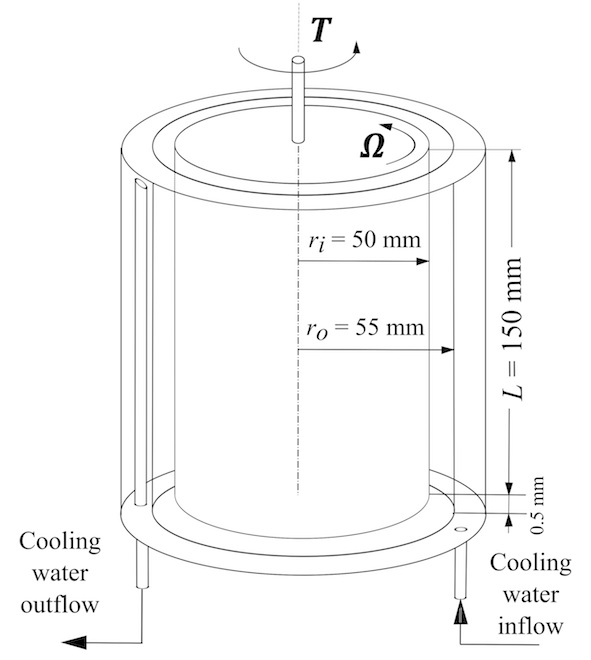} 
	\caption{Sketch of the Taylor-Couette system, drawn to scale.}
	\label{fig1}
\end{figure}

The Taylor-Couette geometry used here is fitted on a rheometer (Physica MCR 501, Anton Paar). Figure \ref{fig1} shows a sketch of the experiment. The inner cylinder has a radius $r{_i}=50\pm0.01~mm$ and the outer cylinder $r{_o}=55\pm0.01~mm$. Hence, the gap between both cylinder is $d=r{_o}-r{_i}=5\pm0.01~mm$. The length of the inner cylinder is $L=150\pm0.5~mm$. Consequently, the dimensionless parameters which describe the geometry are the radius ratio $\eta=r_i/r_o=0.909$ and the aspect ratio $\Gamma=L/d=30$. The object of the present experiments was to examine the relationship between the angular speed of the inner cylinder,  $\Omega$, and the torque that it exerts on the fluid,  $T$. The Reynolds number, \textit{Re}, is based on the angular velocity of the inner cylinder, the inner cylinder radius, the gap between cylinders and the properties of the fluid: $Re=\Omega r_i d/\nu$, where $\nu$ is the kinematic viscosity of the working fluid. The dimensionless torque, $G$, is based on the torque exerted by the fluid in the walls of the inner cylinder, the height of the inner cylinder and the properties of the fluid. $G$ is defined as $G=T/2\pi\rho\nu^2 L$, where $\rho$ is the density of the working fluid. In addition to $G$, the so-called $\omega$-Nusselt number, $Nu_\omega$, defined according to \cite{Eckhardt2007}, is also used:
\begin{equation}
  Nu_\omega = \frac{G}{G_{lam}}, \quad \mbox{where\ }\quad G_{lam}=\frac{2 \eta}{\left(1+\eta\right)\left( 1-\eta \right)^2}Re,
  \label{Nuomega}
\end{equation}
$Nu_\omega$ represents the torque measured in the units of laminar torque.

The inner cylinder is made of aluminum and its surface is anodized. The bottom of the inner cylinder is recessed. There is a gap  of $0.5\pm0.001~ mm$ between the edge of the base of the inner cylinder and the flat bottom of the outer cylinder filled with an air bubble, which minimises the shear stress on the bottom of the cylinder. The top part of the gap between both cylinders is covered with an annular PVC lid. It is positioned so that the bottom of the lid is at the same height as the upper edge of the inner cylinder. That makes the gap being completely filled and there is no contact between the lid and the inner cylinder. The outer cylinder is made of glass and there is an additional glass jacket connected to a flow of water in order to maintain the working temperature at $22\pm 0.01^{\circ}$C. For flow visualization purposes, 2\% of Kalliroscope is added to the fluid. The rheometer allows for torque or speed controlled runs. The highest acquisition frequency is 100 Hz and the real resolution of the encoder is smaller than $1~\mu rad$. The accuracy of the torque is 0.5\% of the measured value and never smaller than 0.2 $\mu Nm$. Several fluids, mixtures of water and glycerol, were used in order to optimize the speed acquisition. Additional measurements were done using a low viscosity silicon oil in order to reach data in the highest range of {\it Re}.

\subsection{Experimental procedure}
In quasi-static ramping  of the velocity, the Taylor vortex flow is characterised by 30 time-independent, axisymmetric, toroidal vortices from $Re_c=138$. This value is close to the predicted value (139) given by the stability theory \citep{Esser1996}. For $157<Re<199$, wavy vortex flow is seen. Then, from $Re=199$, modulated wavy vortex is observed. 

\begin{figure}
	\centering
	\includegraphics[width=0.95\columnwidth]{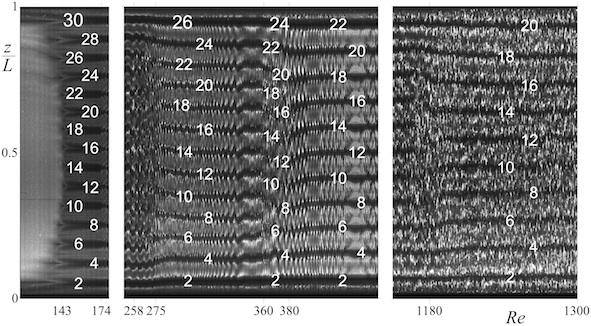} 
	\caption{Spatio-temporal diagram over the whole height $\left(\Gamma=L/d=30\right)$ of the Taylor-Couette flow $\left(\eta=0.909\right)$ in three intervals from $Re=110$ to 1300. The acceleration rate is : $\partial Re / \partial \tau = 4$. The numbers on the diagram count the number of cells from bottom to top.}
	\label{fig2}
\end{figure}

In figure \ref{fig2}, a spatio-temporal diagram over the whole height of the cylinders is presented. The flow is driven by the rotation of the inner cylinder at a constant ramping rate: $\partial Re / \partial \tau = 4$, where $\tau = t \nu / d^2$ and $t$ is time. As \textit{Re} increases from the laminar Couette flow, Eckman vortices develop at the ends of the cylinders \citep{Czarny2003}. Then these vortices evolve into a well defined state of axisymmetric steady toroidal Taylor vortices that rapidly join at the center of the cylinder at $Re=143$. At $Re=174$, the wavy vortex flow starts. In figure \ref{fig2}, the second interval ($250<Re<450$) shows sequences of regions of strong modulation leading to the merging of cells (26, 24 and 22). In this same range of \textit{Re}, \cite{Coles1965} showed in a system of similar aspect ratio that a large number of expected states were accessible due to the vertical oscillations and the merging of these cells. Clearly, as we progress in time or in \textit{Re} the number of cells decreases from 30 to 20. To obtain a smaller number of cells, say 18, a lower acceleration is used. Hence, this merging of cells will allow to prepare an initial state with 30 to 18 cells. Only the number of cells is taken into account and spiral modes are not considered here. Once the desired number of cells is set up, our strategy is to change instantaneously the torque to a prescribed value and then measure the velocity for few minutes before the next measurement. The stability of each state was tested during approximately 70 times the viscous time, {\it i.e.} $\nu/d^2$, to ensure that the number of vortices remained constant. The velocity fluctuations of the rotating cylinder are small, typically around 0.6\%.  

\section{Results and discussion}
The results are a combination of torque measurements and simultaneous flow visualisations. The first set of results is concerned with the properties of two distinct states: 30 and 18 cells states. The second set of results presents and analyses the torque data, describes the stability limit of the states as well as the change in behaviour in the torque-speed curves for different states.

\subsection{Aspect ratio of cells}%
\begin{figure}
\begin{center}
	\includegraphics[width=0.9\columnwidth]{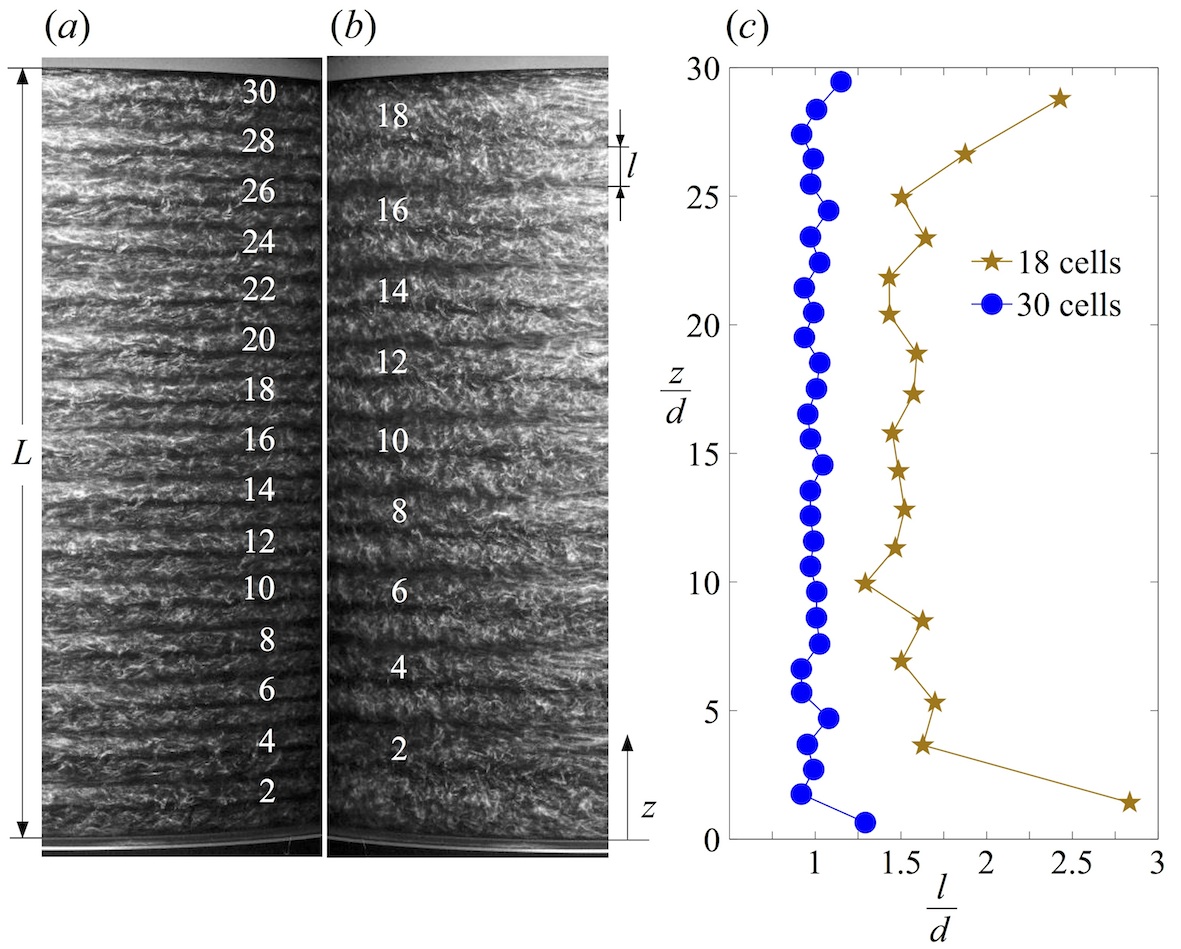}
	\end{center}
\caption{Photographs and aspect ratio along the axis of 30 and 18 cellular modes at $Re=6000$. ({\it a}) Normal 30 cells, ({\it b}) abnormal 18 cells, and ({\it c}) the associated aspect ratio along the axis.}
\label{fig3}
\end{figure}

The finding of different states for the same boundary conditions requires that the size of the cells varies from one state to the other. In figure \ref{fig3}({\it a}) and ({\it b}), photographs of two states with 30 and 18 cells at $Re=6000$ are presented. Figure \ref{fig3}({\it c}) presents their aspect ratio along the vertical axis. For intermediate number of cells: 20, 22, 24, 26 and 28, intermediate curves are expected. For the primary state of 30 cells, the aspect ratio, $l/d$, of all the cells is close to one. However for the 18 cells pattern, the ratio $l/d$ is 1.5 in the center of the cylinder so the cells are elongated. Note the cells close to the ends of the cylinders have a significant larger aspect ratio up to 2.5. This indicates that the caps have a local effect and strongly elongate the two cells close to the ends \citep[]{Czarny2003}. Although the data on the aspect ratio are for $Re=6000$, using the same protocol, we have observed states with different numbers of vortices over a wide range of {\it Re}, that looks essentially similar to the ones in figure \ref{fig3}({\it a},{\it b}).

The states in figure \ref{fig3} were obtained at $Re=6000$. The difference in the measured torque between the 18 cells state and the 30 cells state is 13.5\%. The state with the larger number of cells experiences a larger torque. This is in agreement with the numerical simulation of \cite{Ostilla2013} although their range of \textit{Re} $\left(400<Re<1600\right)$ and the number of cells tested (6 or 8 cells)  is smaller.

\subsection{Torque}%
\begin{figure}
	\centering
	\includegraphics[width=0.8\columnwidth]{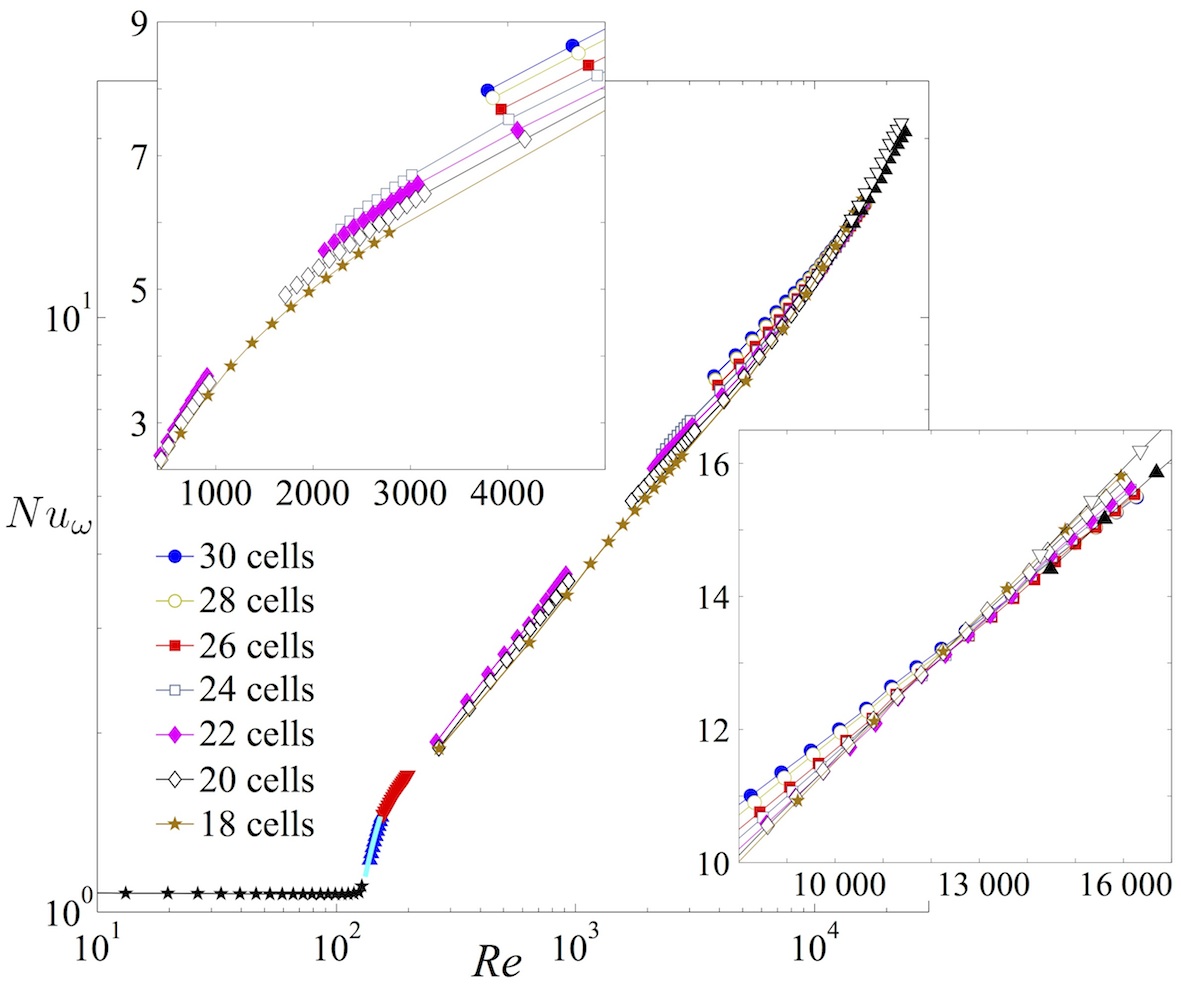} 
	\caption{Rescaled torque, $Nu_\omega$, as a function of \textit{Re} for the different flow states. The black stars ($\bigstar$), the blue triangles (\textcolor{blue}{$\blacktriangle$}) and the red triangles  (\textcolor{red}{$\blacktriangledown$}) represent the Couette flow, the Taylor vortex flow and the wavy vortex flow, respectively. The thick (cyan) line represents a fitting using the proposed scaling of \cite{Donnelly1960}. Top inset is a zoom at small \textit{Re} in linear scale showing the stability limit of some states. Bottom inset is a zoom at large \textit{Re} in linear scale showing the intersections. The black triangles ($\blacktriangle$) and the empty triangles ($\triangledown$) represent data obtained using low viscosity silicone oil. The error bars are smaller than the plotting symbols.}
	\label{fig4}
\end{figure}

Torque and speed measurements have been performed up to a \textit{Re} of 24 000. Using the procedure presented earlier we were able to measure the torque associated to 7 different states: 30, 28, 26, 24, 22, 20 and 18 cells. 
The relationship between the $\omega$-Nusselt number and the Reynolds number for these states is presented in figure \ref{fig4}. Different symbols and colors represent the different states: the laminar Couette flow, the Taylor vortex flow (30 cells), the wavy vortex flow and the 7 different turbulent Taylor vortex flows with 30, 28, 26, 24, 22, 20 and 18 cells. In order to access large values of  {\it Re} up to 24 000, low viscosity silicone oil was used and two series of data are reported in figure \ref{fig4}. The use of silicone oil does not allow to visualise the flow and therefore to count the number of vortices.  These data were obtained by applying protocols leading to states with large and small number of cells. The trend of the curves suggests that the black triangles correspond to large number of cells and empty triangles to small number of cells.

In figure \ref{fig4}, the $\omega$-Nusselt number associated to Couette flow ($Re<Re_c$) is almost constant and just above one; the small difference from one is due to end effects. The Taylor vortex flow and the wavy vortex flow have distinct properties, corresponding to different slopes in the $Nu_\omega$ versus {\it Re} curve. The rescaled torque behaviour for Taylor vortex flow ($138<Re<154$) is fitted by a relation suggested from the finite amplitude theory by \cite{Stuart1958} and proposed by \cite{Donnelly1960}:
\begin{equation}
  Nu_\omega = a Re^{-2} + b Re^{0.36},
  \label{DonnellySimonEq}
\end{equation}
with $a=-13~374$ and $b=0.33$.

The stability domain of the different number of vortices is in the top inset of figure \ref{fig4}. It is interesting to notice that the 30, the 28 and the 26 vortices states are not stable for \textit{Re} below 3800. Similarly, the 24, the 22 and the 20 cells states are not stable for \textit{Re} below 2300, 2100 and 1700, respectively. In the range between 1000 and 1700, only the 18 cells state is stable. These results are reminiscent of the stability studies from \citet{Coles1965}, \cite{Snyder1969},  \cite{Koschmieder1979} and \cite{Cliffe1992}. 

As the Reynolds number increases, a systematic increase of $\omega$-Nusselt is observed. The 18 cells state has always the lowest $\omega$-Nusselt whereas the states associated to the largest number of cells have always the highest $\omega$-Nusselt. As the Reynolds number increases further the curves $Nu_\omega$ versus $Re$ come closer. Then, the curves for different states intersect in a range of {\it Re} between 9600 and 15 500. A zoom in this range is shown in the bottom inset of figure \ref{fig4} where most of the intersections are located around 12 600. Below each intersection, the torque is larger for states where the number of cells is larger. Above each intersection, the trend changes and a smaller number of cells leads to larger torque. Similar intersections were reported by \cite{Lathrop1992} and \cite{Lewis1999} for eight and ten cells states in a system with a radius ratio of 0.724.

\begin{figure}
	\centering
	\includegraphics[width=0.7\columnwidth]{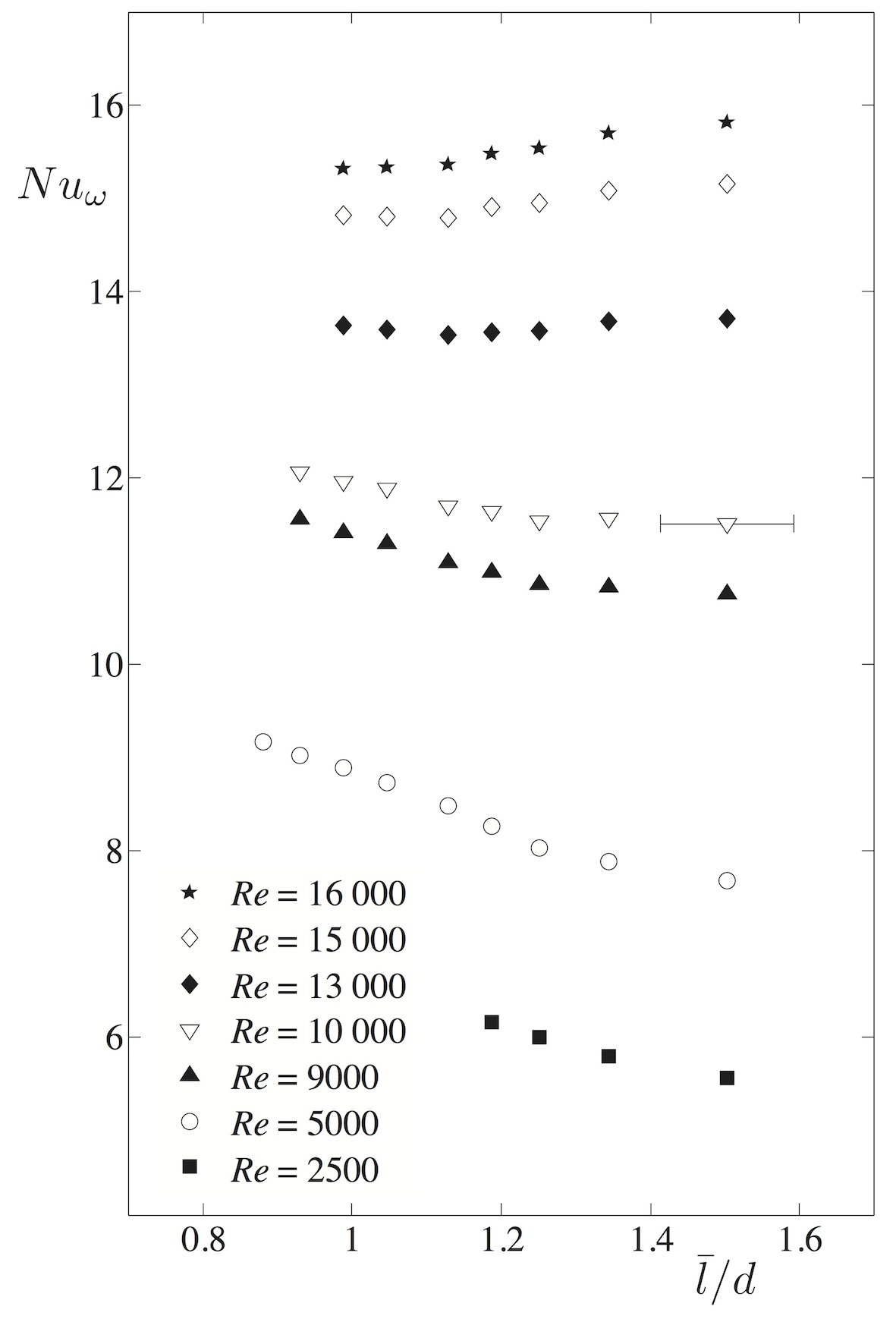} 
	\caption{The rescaled torque in dependence on the aspect ratio of the cells for different \textit{Re}. The horizontal error bar for $Re=10~000$ represents the maximum error on $\Bar{l}$. The vertical error on $Nu_\omega$ is smaller than the symbol height. }
	\label{fig5}
\end{figure}

The analysis of our data allows to present, in figure \ref{fig5}, the rescaled torque as a function of the aspect ratio of the vortices, $\Bar{l}/d$, for several values of $Re$. For relatively small $Re$, the rescaled torque decreases as the aspect ratio of the vortices increases. For large $Re$, the rescaled torque increases with  $\Bar{l}/d$. At $Re=13~000$, $Nu_\omega$ is almost constant. In the experiments, $\Bar{l}/d$ between 0.88 to 1.5 correspond to states with 34 to 18 cells. Our results confirm a clear effect of the vortex aspect ratio on $Nu_\omega$, which was predicted by numerical simulations \citep{Brauckmann2013,Ostilla2013}. For $Re=5000$, our values of $Nu_\omega$ are slightly lower than the numerical results of \citet{Brauckmann2013} and do not exhibit a peak that these authors found  in their data at  $\Bar{l}/d=0.965$. The discrepancy between  the experiments and the numerical simulations may be due to the fact that the numerical simulations were performed for a radius ratio of 0.71.

\subsection{Exponent}%

The dependence of $Nu_\omega$ on \textit{Re} can be described by 
\begin{equation}
  Nu_\omega=\mathcal{A} ~ Re^{\alpha-1},
  \label{Eckhardt}
\end{equation}
where $\mathcal{A}$ depends on the vortex aspect ratio and is decreasing function of {\it Re}. A motivation behind seeking scaling laws is that insight into the turbulence mechanisms can be uncovered \citep[see][]{Dubrulle2005,Eckhardt2007}. It is clear that the scaling laws discussed above cannot hold for Taylor vortex flow and wavy vortex flow since the flow is non-turbulent. \cite{Donnelly1960} discussed this issue and proposed the scaling (\ref{DonnellySimonEq}) for Taylor vortex flow. For the range of \textit{Re} below 3000, the exponent found in the present study is almost constant ($\alpha\simeq1.5$) and agrees well with the data from \cite{Lim2004} and \cite{Ravelet2010}.

\begin{figure}
	\centering
	\includegraphics[width=0.95\columnwidth]{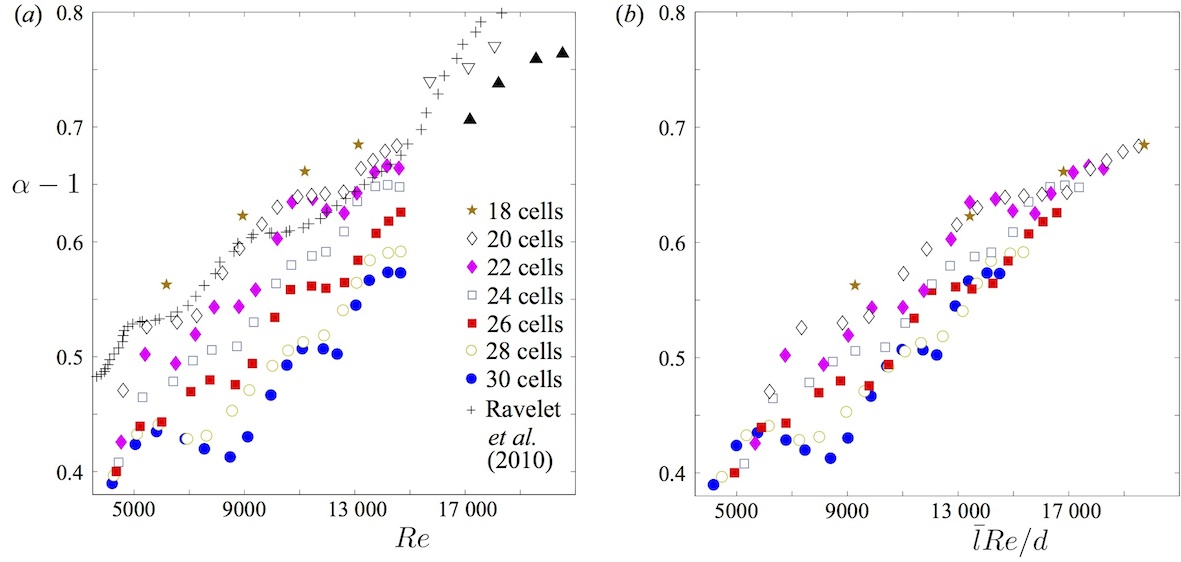} 
	\caption{Evolution of the exponent for different number of cells. The black triangles ($\blacktriangle$) and the empty triangles ($\triangledown$) represent data obtained using low viscosity silicone oil. ({\it a}) $\alpha-1$ as a function of {\textit Re}  and ({\it b}) $\alpha-1$ versus $\Bar{l} Re/{d}$, based on the averaged height of the cells,  $\Bar{l}$}
	\label{fig6}
\end{figure}

Figure \ref{fig6}({\it a}) displays the variation of $\alpha-1$ as a function of \textit{Re} for different number of cells for  $4000\lesssim Re\lesssim21~000$. Here, $\alpha-1=\partial\left(\log_{10}Nu_\omega\right)/\partial\left(\log_{10}Re\right)$ is calculated for each state separately as a function of the \textit{Re} using a sliding least square fit \citep{Lathrop1992,Lewis1999,Ravelet2010,Merbold2013} over the interval $\Delta\left(\log_{10}Re\right)=\Delta_{10}=0.1$. For the 18 cells state the interval is 0.2. The results are compared with \cite{Ravelet2010} whose experiment has $\eta=0.917$ and $\Gamma=22$. Although, their aspect ratio is smaller then ours, there is a good agreement between their results and the present exponents for the 20 and 22 cells states. Both studies exhibit monotonous increase for $3000\lesssim Re\lesssim15~000$. 

A new scaling of the Reynolds number based on the mean aspect ratio of the cells, $\Bar{l}Re/d$, is proposed for the exponent in figure \ref{fig6}({\it b}). Only a partial collapse of the data is found. This suggests that the size of the cells is an important parameter in the turbulent regime studied here. Note that at higher $\Bar{l}Re/d$, the collapse is better. 

The meaning of $\alpha$ is related to the viscosity dependence of the torque \citep[]{Lathrop1992}. The Kolmogorov assumption assumes $\alpha=2$ for fully developed turbulence \citep{Doering1992}. Any deviation from $\alpha=2$ implies a particular form of velocity fluctuations. In the flow system with a given number of vortices, the form of velocity fluctuations is constrained to the presence of large scale vortices and cannot be completely random. Modern investigations using particle image velocimetry \citep[see][]{Racina2006,Tokgoz2012} aim at quantifying the velocity fluctuations and estimating the average turbulent kinetic energy dissipation rate.

The principal feature, we wish to highlight, is the systematic dependence of the scaling on the aspect ratio of the cells, displayed in figure \ref{fig6}. For situations where the number of cells is small, typically 18 here, the flow is less  constrained by the cells. Therefore, the turbulent flow will exhibit larger velocity fluctuations leading to higher turbulence and a larger $\alpha$. For situations were the number of cells is large, typically 30, the flow is more constrained by the cells and the velocity fluctuations are weaker leading to a smaller $\alpha$. 

\section{Conclusions}

The effect of the number of vortices on torque in a Taylor-Couette flow ($\eta=0.909$ and $\Gamma=30$) up to $Re=24~000$ has been quantified. A specific protocol was used to obtain different states with 34, 32, 30, 28, 26, 24, 22, 20 and 18 cells. The evolution of the rescaled  torque, $Nu_\omega$, versus \textit{Re} for different number of cells is obtained. The results are in agreement with \cite{Lim2004} and \cite{Ravelet2010} and, moreover, the curves superpose self-consistently. 

The effect of the vortex size on the rescaled torque indicates a change in behaviour corresponding to the intersection in the $Nu_\omega$ versus {\it Re} curves at {\it Re} between 9600 and 15 500. For most of the states, the intersection is around 12 600. Before each intersection, the torque is larger for large number of cells and smaller after the intersection. In the same range of {\it Re}, but for a system with radius ratio of 0.725, \cite{Lathrop1992} found a transition from centrifugal instability to shear turbulence using torque and local wall shear stress data. \cite{Lewis1999} confirmed this transition by means of additional data of torque and wall shear stress, as well as velocity measurements. The data from \cite{Lewis1999} shows a similar intersection for eight and ten cells states. In our case, it is not possible to conclude that the intersection is an indicator of the transition to shear-driven turbulence.

The scaling exponent of the torque is larger for states with large aspect ratio. It is also found that the exponents collapse when scaled with a Reynolds number based on the aspect ratio of the vortices.  Finally, new large experimental apparatuses have been built \citep{vanGils2011,Merbold2013,Avila2013} and these effects will be further detailed in ranges of higher \textit{Re} if the future investigations dare count the number of vortices.

\begin{acknowledgments}
We are grateful to Florent Ravelet for providing the data from \cite{Ravelet2010} and Sebastian Merbold for discussions on the calculation of the exponents. The authors also acknowledge the financial support of the R\'egion Haute Normandie and the Agence Nationale de la Recherche (ANR), through the program ``Investissement d'Avenir'' (ANR-10-LABX-09-01), LabEx EMC3. 
\end{acknowledgments}

\bibliography{TCjfm}
\bibliographystyle{jfm}

\end{document}